**Interfacial strength dominates fold formation in microscale, soft static friction**

Justin D. Glover[1†], Xingwei Yang[2†], Rong Long[2], Jonathan T. Pham[1*]

[1] Department of Chemical and Materials Engineering, University of Kentucky, Lexington, KY 40506, USA

[2] Department of Mechanical Engineering, University of Colorado, Boulder, CO 80309, USA

†Equal contribution. *Email: Jonathan.Pham@uky.edu.

**Abstract**

Utilizing colloidal probe, lateral force microscopy and simultaneous confocal microscopy, combined with finite element analysis, we investigate the mechanism of static friction for a microparticle on a soft, adhesive surface. We find that the surface can form a self-contacting fold at the leading front, which results from a buildup of compressive stress. A sufficiently high lateral resistance is required for folding to occur, although the folds themselves do not increase the peak force. Experimentally, folds are observed on substrates that exhibit both high and low normal adhesion, motivating the use of simulations to consider the role of adhesion energy and interfacial strength. Our simulations illustrate that the interfacial strength plays a dominating role in the formation of folds, rather than the overall adhesion energy. These results reveal that adhesion energy alone is not sufficient to predict the nucleation of folds or the lateral force, but instead the specific parameters that define the adhesion energy must be considered.

**Introduction**

Investigation of friction started half a millennium ago with Leonardo da Vinci,[1] yet challenges in studying friction persist today, especially for soft materials.[2,3] An understanding of soft friction is important for many applications, from bioinspired adhesives[4-6] and car tires[7] to damage of articular cartilage.[8] In the most conventional form, one can consider a rubber block attached to a spring, which is pulled on a rigid surface. The block remains stationary until the force overcomes a static threshold, at which point the block moves along the surface. Static friction, defined as the lateral resistance between two surfaces that are not yet in relative motion, is related to the normal loading on the block and the true contact area at the interface.[4,9-11] Since most surfaces possess microscale roughness, microscopic contact points usually control the true contact area.[12,13] Accordingly, it is generally accepted that microscale contacts govern static friction behavior observed on the macroscale; however, the mechanism by which a single microscopic point starts to move along a soft adhesive surface has not been well explored.

To study friction and adhesion of a single contact, a spherical probe is commonly used as a model.[9,13-19] When a spherical probe is pulled laterally on a soft elastomer, static friction is observed due to adhesive interactions between the sphere and the substrate.[20] In this situation, Schallamach waves often emerge as the mechanism to enable relative motion. First reported in 1971, Schallamach waves describe a cascade of detachment and reattachment events that propagates through the contact zone.[21,22] These waves form when a buckling event occurs at the moving front, which then reattaches to the probe to create a pocket of air. This air gap then moves through the contact zone, like a ruck moving through a rug,[23] without significant interfacial slippage between the surfaces. However, the governing parameters that define Schallamach waves are still unsolved, as they change depending on the loading and surface conditions.[24] Furthermore, spherical probe experiments have focused mainly on the macroscale,[22,25-30] likely due to challenges in visualizing and manipulating small contacts; however, dimensional analysis suggests that the effect of adhesion should become more prominent as the length scale of contact approaches the elasto-adhesive length.[31,32] Even more broadly, the relationships among adhesion energy, interfacial strength, and static friction remains a perplexing question. For example, the peel force of an adhesive film has been shown to trend inversely with surface energies, which is a non-intuitive finding.[33] Hence, a surprising gap in our knowledge still exists on soft static friction, especially for small scale contacts.

Here, we focus on how a stiff microparticle starts moving laterally on a soft adhesive surface by combining confocal microscopy, colloidal probe, lateral force microscopy, and numerical simulations. Our results show that a self-contacting fold can form at the leading front. Nucleation of folds is the result of strong interfacial interactions between the particle and the substrate, although the fold itself does not necessarily raise the peak lateral force. Intriguingly, our experimental results illustrate that fold formation can occur for situations of either low or high normal adhesion. This motivates simulations that consider the role of adhesion energy versus interfacial strength on the emergence of folds. Our simulations provide compelling evidence that suggests interfacial strength, rather than adhesion energy, is the governing factor for the formation of folds.

**Results and Discussion**

*Experimental approach and initial observations*

For our experiments, we use a colloidal probe on an atomic force microscope (AFM), combined with a high-precision stage, to apply lateral displacements and measure lateral forces on a soft, polydimethylsiloxane (PDMS) elastomer (Fig. 1a). To image the substrate, a confocal microscope with a piezo-driven objective is used for fast, *in-situ* imaging of x-z cross-sections. For the PDMS substrate, Sylgard 184 is mixed at a 60:1 base:crosslinker ratio and spin-coated on a glass slide to a thickness of ~90 μm. Prior to spin-coating, a fluorescent monomer is added that binds to the polymer network, enabling fluorescent imaging by confocal microscopy.[34-36] Upon curing, a crosslinked film is obtained with a Young's modulus of approximately 4 kPa.[34,37,38] For the colloidal probe, an 8.5 μm radius (*R*) glass sphere is attached to an AFM cantilever.

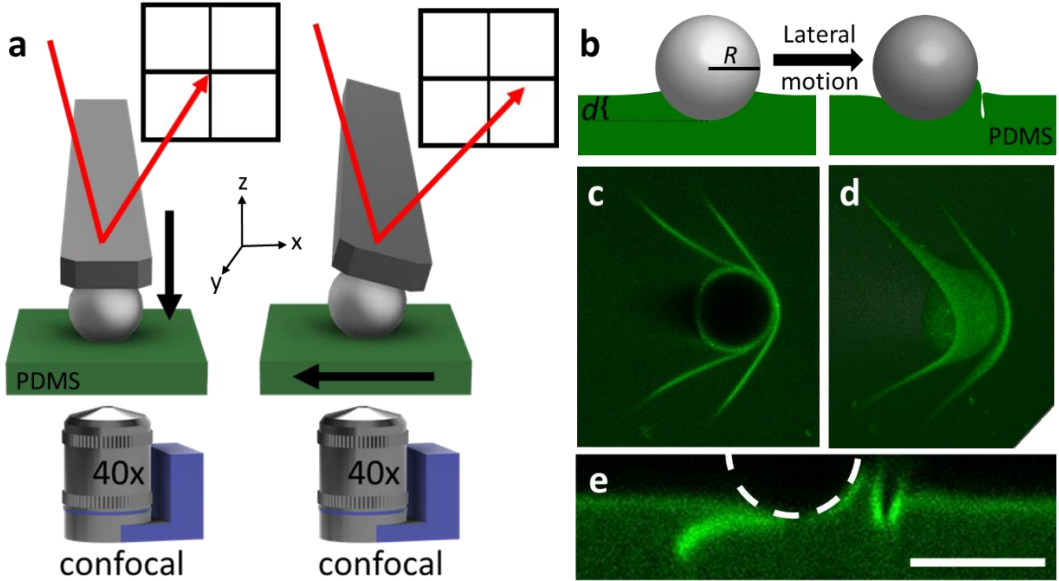

**Figure 1. Experimental observations of a microparticle starting to be pulled laterally on a soft elastomer.** a) Schematic of the experimental setup showing the combined AFM cantilever and confocal microscope used to simultaneously image the contact and measure the lateral force. b) Schematic of fold formation, also showing the experimental variables of depth, $d$, and radius, $R$. c) An x-y confocal image of a microsphere ($R = 8.5$ μm) with two folds; one fold is under and the other is in front of the microsphere. d) 3D image of the same microsphere viewed from below. Note that the particle is stopped for high resolution 3D images. e) An x-z cross-sectional image showing two folds. Scale bar: 20 μm.

In a typical experiment, the sphere is brought into contact with the surface to a desired depth, $d$, and held for 15-20 seconds. The substrate is subsequently translated laterally at a pre-defined velocity, $v$, while simultaneously imaging the contact (Fig. 1a). To measure the lateral force, $F_L$, a laser bounces off the cantilever into a detector to quantify cantilever deflections. As the PDMS is pulled laterally relative to the stationary sphere, the cantilever deflects and changes the laser position (Fig. 1a). To investigate the effect of indentation depth, the relative depth is controlled to a range of $d/R$ ~ 0.03 to 0.9. The upper limit is chosen such that the substrate meniscus remains at or below the center of the sphere prior to translation, while the lower limit is set naturally by adhesion, which slightly pulls the sphere into the substrate. Since rate has been shown to affect friction behavior, three velocities are chosen over a few decades: $v =$ 10, 1, and 0.1 µm/s.[17,39]

Upon translation, the PDMS substrate can fold at the leading front (Fig. 1b). As the substrate continues to translate, a second fold also forms before the first fold releases. Example confocal images of folds are shown in Figs. 1c-1e from different viewing directions, depicting a full picture of the fold geometry. As illustrated in a top-view x-y image (Fig. 1c), the fold extends outside the contact zone of the microsphere. A 3D bottom-view (Fig. 1d) shows that the first fold stretches underneath the microsphere, while a second fold forms at the leading edge. A cross-sectional x-z image through the microsphere centerline, directly before a fold collapses (Fig. 1e), clearly illustrates the geometry of a fold. Although we find this folding process to be the most interesting, the surface does not always fold upon translating the substrate. Videos of the two cases (folding and non-folding) are presented in the Supplementary information (Supplementary videos 1 and 2), and snapshots of important events are provided in Figs. 2a and 2b. To understand what causes the two different cases on seemingly similar materials, we perform several control experiments (see Supplementary Note 1) and conclude that aging of the elastomer in ambient light causes folding for our samples; when stored in the dark, folds are not observed. Moreover, freshly prepared, unaged samples also do not exhibit folds. This finding offers an approach to investigate the physical parameters that govern the formation of folds and the corresponding lateral force.

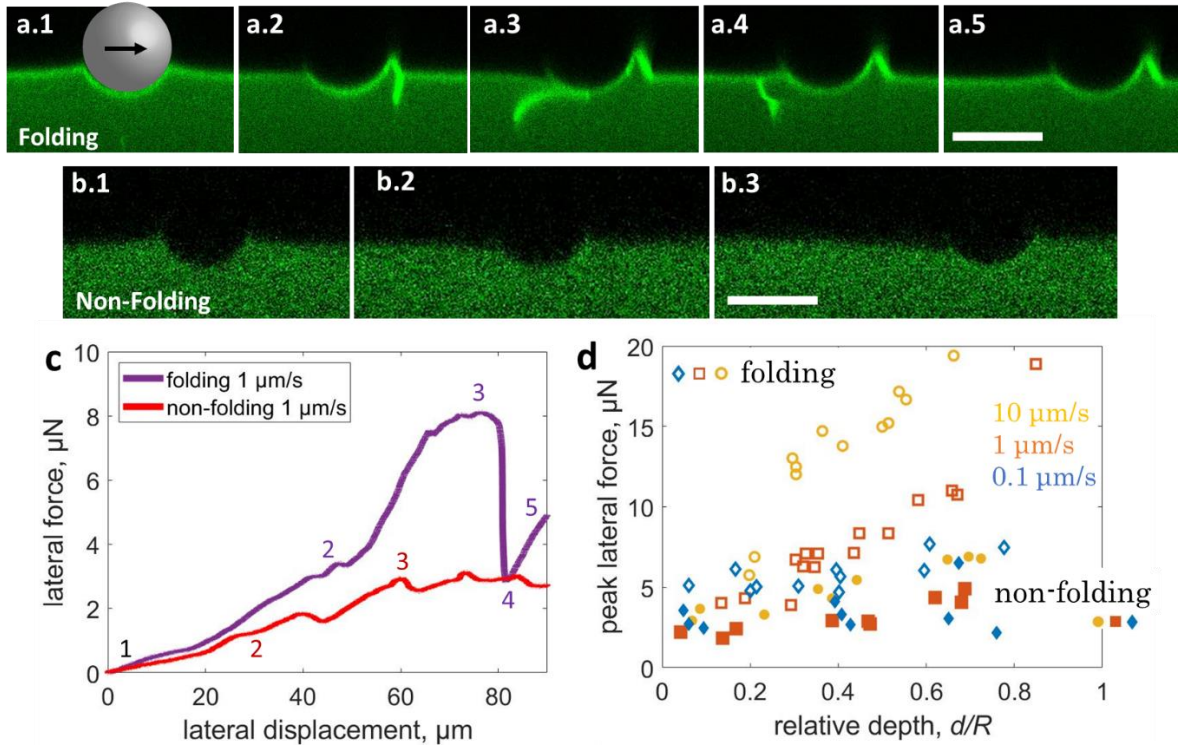

**Figure 2. Folding and non-folding cases and the corresponding lateral forces.** a) Confocal x-z images of a folding sample moving at $v$=1 μm/s with $d/R$=0.4. a.1). The microsphere in contact with the surface prior to the start of motion. a.2) Formation of a first fold. a.3) The moment before the first fold is about to release on the trailing edge. a.4) First image frame after the first fold releases and the peak force drops. a.5) Second frame after the first fold releases, showing an unfolded state. b) Confocal x-z images of a non-folding sample moving at $v$=1 μm/s with $d/R$=0.4. b.1) The microsphere in contact with the surface prior to the start of motion. b.2) The contact shape about midway to the peak lateral force ($F_{peak}$). b.3) The contact shape once the lateral force $F_L$ has reached an approximate force plateau. c) Lateral force ($F_L$) vs distance curves for folding and non-folding samples. The numbered points correlate with the confocal images in parts a and b. d) The peak lateral force ($F_{peak}$) for folding and non-folding samples as a function of $d/R$ and $v$. Scale bars: 20 μm.

To quantitatively compare the lateral resistance of the two cases, we use freshly prepared samples and samples aged for 4 days in lit ambient conditions. In Fig. 2c, an example lateral force ($F_L$) versus lateral displacement curve is plotted for folding (purple curve) and non-folding (red curve) cases ($d/R = 0.4$ and $v = 1$ μm/s). With the benefit of simultaneous imaging, $F_L$ is correlated to confocal images to gain insight into the fold-force relationship. The numbered spots on Fig. 2c correlate with the numbers in Figs. 2a and 2b. For example, at spot 1 on Fig. 2c, the microsphere is in contact but at rest on the surface (Figs. 2a.1 and 2b.1). Upon translating the substrate for the folding case, $F_L$ initially increases while the microparticle remains attached to the substrate. At spot 2, a fold emerges, leading to a small reduction in $F_L$ prior to the fold collapsing; this reduction in $F_L$ is afforded by increased compliance from the folded surface. Once the fold is closed, $F_L$ continues to climb and approaches a peak at spot 3. At this point, the first fold is under the microparticle (Fig. 2a.3). Upon translating beyond the peak lateral force (defined as $F_{peak}$), the first fold detaches on the trailing edge (Fig. 2a.4) and $F_L$ drops precipitously to spot 4, returning the surface to an unfolded state (Fig. 2a.5). For the non-folding case, $F_L$ initially increases, like for the folding case. However, a clear difference is observed between the folding and non-folding force plots. At spot 2 (Fig. 2b.2), the height of the PDMS contact line at the back of the microparticle lowers, while no change is evident at the front. After reaching $F_{peak}$, the force is in an approximate plateau region and the most obvious change in deformation geometry is the lower meniscus height at the trailing edge. Some minor stick-slip is observed (Supplementary video 2), which relates to fluctuations in $F_L$.

As demonstrated by Fig. 2c, $F_{peak}$ is significantly higher for folding compared to the non-folding case. To consider whether different testing parameters play a role, $F_{peak}$ is plotted as a function of $d/R$ and $v$ for both cases (Fig. 2d). In general, increasing both $d/R$ and $v$ increase $F_{peak}$ within each case of folding or non-folding. The increase in $F_{peak}$ from $v$ is likely associated with viscoelastic effects, while the increase from $d/R$ is likely due to the increase in contact area. By looking at a single $v$ to compare the two cases (e.g. yellow open vs yellow closed points at 10 μm/s), we find that $F_{peak}$ is higher for folding cases compared to the non-folding cases. On the other hand, the observation of folding versus non-folding between the two types of samples (i.e., aged in ambient light vs freshly prepared) is consistent across two decades of velocities. This implies that the formation of folds is itself not rate-dependent, although the rate can affect the measured forces. These unique experimental observations motivate the following questions:

What governs the formation of folds and how do folds affect the lateral forces? Because the experiments presented in Fig. 2 exhibit higher $F_{Peak}$ for folding samples, it is intuitive to assume that there an increase in PDMS-particle adhesion energy ($W_{ad}$) with light-aging, which leads to the formation of folds.

*Effect of adhesion energy*

To dive deeper into how adhesion is related to the emergence of folds, we employ finite element analysis (FEA) (Supplementary Note 2). The FEA model incorporates the viscoelastic properties of our PDMS, obtained by shear rheology, and the probe diameter is set to be identical to the experiments. Adhesion between the sphere and the substrate is modeled by a cohesive zone with tunable adhesion energy $W_{ad}$. Guided by the experimental data in Fig. 2, we first run simulations with $d/R = 0.4$ and $v=1$ μm/s, and set $W_{ad}$ to either a low (4 mJ/m²) or a high (20 mJ/m²) value. Snapshots in Figs. 3a and 3b are recorded in a similar fashion to the images in Fig. 2. Unlike the experiments, FEA allows for visualization of the horizontal normal strain in the substrate, $\varepsilon_{xx}$, as depicted by the color contours. Fig. 3a.1 and 3b.1 show the sphere prior to the start of motion. As the sphere starts to move laterally in the high $W_{ad}$ simulation, a fold forms in front of the sphere where $\varepsilon_{xx}$ is highly compressive (Fig. 3a.2). Matching the experiments, the fold grows in size until it detaches (Fig. 3a.3 and 3a.4). The trailing edge features a region with high tensile strain prior to detachment (Fig. 3a.3). After detaching, the fold opens due to the high tension (Fig. 3a.4), and the surface returns to a low strain state while a new fold starts to form (Fig. 3a.5). However, for the low $W_{ad}$ case, no folds form (Fig. 3b.2) and the substrate displays an asymmetric deformation, similar to that observed in our non-folding experiments (Fig. 2b). The absence of folds in the low $W_{ad}$ case is consistent with the significantly lower compressive strain $\varepsilon_{xx}$ in front of the sphere, compared to the high $W_{ad}$ case. Moreover, no folds are apparent at $F_{peak}$ (Fig. 3b.3), demonstrating that the sphere slips before a fold is able to form. A plot of $F_L$ versus the translating lateral displacement (Fig. 3c) confirms that $F_{peak}$ is higher for the folding case. Also consistent with our experiments, simulations show an increase in $F_{peak}$ with $d/R$ and $v$ (Fig. 3d).

Despite the qualitative agreement, quantitative discrepancies exist between the FEA and experimental results that should be acknowledged. For example, the lateral forces obtained in the FEA (Fig. 3d) are a few times smaller than the experimental data (Fig. 2d). Additionally, in the non-folding case, FEA suggests that the lateral force drops to a low plateau after $F_{peak}$ (Fig. 3c), while experiments show that the plateau is at a level similar to $F_{peak}$ (Fig. 2c). These discrepancies are attributed to a number of assumptions adopted to make the FEA model tractable while capturing the essential physics. Further discussions on these assumptions are detailed in Supplementary Note 3. Nevertheless, FEA provides the following physical picture for fold formation: Under high $W_{ad}$, a highly compressive region materializes at the leading edge, which triggers the formation of a self-contacting fold through a creasing type of instability. Although these FEA results offer insight into the role of adhesion energy on the folding behavior, they have yet to be experimentally tested.

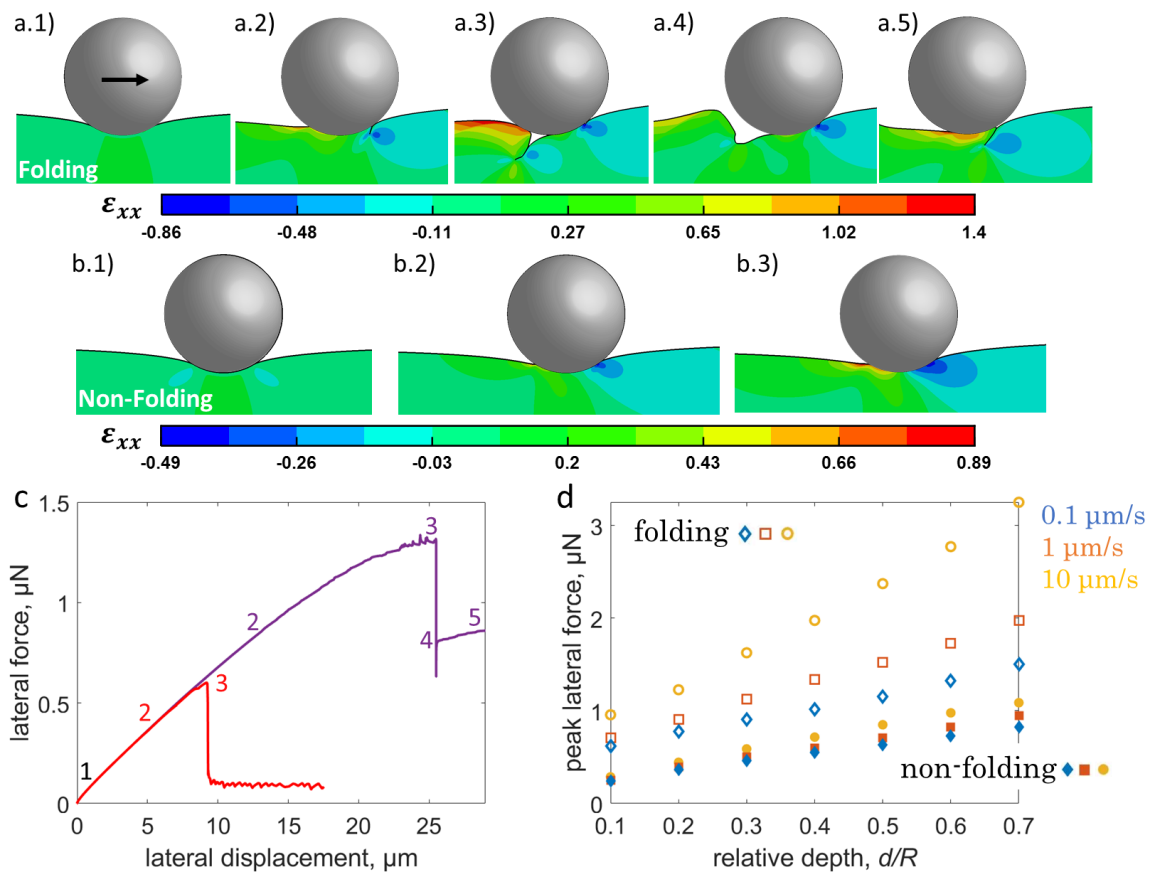

**Figure 3. Capturing the folding and non-folding cases in FEA simulations.** a) A 2D simulation with $v$ =1 μm/s, $d/R$ =0.4, and $W_{ad}$ =20 mJ/m², in comparison to experiments in Fig. 2. The logarithmic normal strain in the x direction $\varepsilon_{xx}$ is shown at different states. a.1) The microsphere in contact with the surface prior to the start of motion. a.2) Initiation of folding. a.3) The moment before the first fold releases, which is at $F_{peak}$. a.4) The moment after the first fold releases from the trailing edge and the peak force drops. a.5) The state after the fold has released while the next fold starts to form. b) A simulation with the same $v$ and $d/R$ but with $W_{ad}$=4 mJ/m²; $\varepsilon_{xx}$ is shown in different states. b.1) The microsphere in contact with the surface prior to lateral motion. b.2) The contact about midway to $F_{peak}$. b.3) The contact once it reaches $F_{peak}$. The color bars for a and b are drawn below each part. c) Lateral force vs lateral displacement curves for folding and non-folding simulations; the number points correlate with the snapshots in parts a and b. d) Simulation data of $F_{peak}$ for folding and non-folding cases as a function of $d/R$ and $v$.

*Interfacial strength governs folding: Comparing experiments to FEA*

With simulations showing that higher $W_{ad}$ leads to the formation of folds, we set out to test this prediction experimentally. One way to increase adhesion of PDMS to glass is by short-duration UV-ozone treatment (UVO), which we anticipated to be easier to control than ambient light-aging.[40] Therefore, freshly prepared PDMS samples are UVO-treated from 0 to 40 s; this range is expected to have a surface modification depth of the order ~1 nm,[41] as opposed to longer UVO treatments (~ 10 minutes or more) that create a thicker glassy layer.[42] Upon lateral testing at $d/R = 0.1$ and $v$=1 μm/s, samples treated for 10 s or less show no folds (Fig. 4a). For samples treated for 20 s or more, folds appear and increase in size with increasing exposure. Fold size is defined here as the length of the self-contacting fold before it is pulled underneath the probe. To compare UVO-treated surfaces with light-aged surfaces, we conducted the same set of experiments but with aging in ambient light conditions over the course of 1 to 4 days. Folds appear on light-aged samples after 1 day and continue to grow in size up to 4 days (Fig. 4b) (recall the data in Fig. 2 is light-aged for 4 days). The folding behavior in the UVO and light-aged surfaces are similar, with folds that grow with increasing exposure. For comparison, a fold after 40 s UVO (Fig. 4c) appears nearly identical to a fold after 3 days of light-aging (Fig. 4d). In addition, increasing exposure time in both cases initially increases $F_{peak}$; however, a dip occurs once the fold size reaches approximately ~4 µm (i.e. 30 s UVO and 3-day light aging). This non-monotonic behavior suggests an intricate coupling between folding and lateral force: Although nucleation of folds requires a high lateral force, further growth of the folds can lower $F_{peak}$.

Since folds occur on both UVO and light-aged samples, we sought to confirm an increase in adhesion energy for both cases. To measure adhesion, normal indentation and pull-off tests are employed with the same colloidal probe (Fig. 4e). According to the Johnson-Kendall-Roberts (JKR) theory),[21] the pull-off force is proportional to the adhesion energy $W_{ad}$. Unfortunately, the freshly prepared samples and UVO-treated samples display a pull-off distance outside the 15 $\mu m$ vertical limit of our AFM. However, the force-height data are at least consistent with the expectation of increasing normal adhesion. Surprisingly however, the pull-off distance and the work of separation (area under curve) decrease with increased aging for the light-aged samples (Fig. 4f). Curiously, this means that the role of $W_{ad}$ for the light-aged samples is not identical to UVO-treated surfaces, signaling that a generic adhesion energy cannot be the only factor controlling the nucleation of folds and the resulting lateral forces.

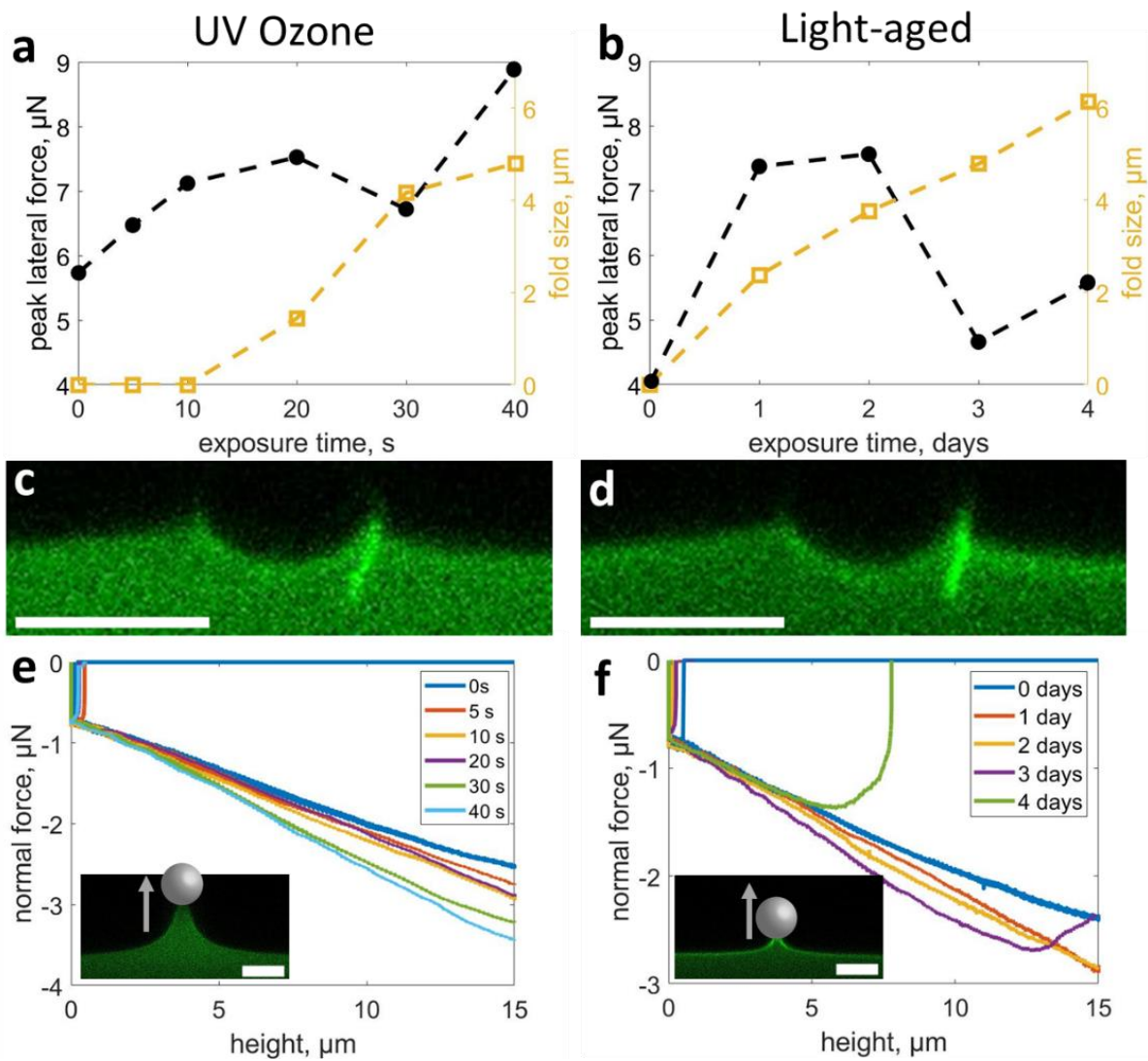

**Figure 4. Peak lateral force, fold size, and normal adhesion with surface treatments.** A) Fold size and $F_{peak}$ as a function of increasing UV ozone time. B) Fold size and $F_{peak}$ as a function of increasing time in ambient light conditions. C) A fold on a 40 s UVO treated surface. D) A fold on a 3-day light-aged surface. E) Vertical adhesion force vs. height curves of a surface exposed to UVO for different times. Inset: Confocal image showing maximum vertical extension for a 40 s UVO treated surface. F) Vertical adhesion force vs. height curves of a sample with different light-aging times. Inset: Confocal image of maximum vertical extension for a 4-day light-aged sample. Scale bars: 20 µm.

Since the folds theoretically originate from a creasing instability,[43] their nucleation should be governed by the maximum compressive strain attainable in front of the sphere; this maximum strain scales with the maximum adhesive traction exerted on the elastomer. The adhesion energy $W_{ad}$, defined as the energy required to separate a unit area of interface, is related to but not equivalent to the adhesive traction. In our simulations, adhesion is described by a bilinear function relating the magnitude of adhesive traction $\sigma$, and the relative separation $\delta$, between two points that are initially in contact (Fig. 5a). Hence, $W_{ad}$ is equal to the area underneath the triangle defined by $\sigma(\delta)$, i.e., $\sigma_{max}\delta_f/2$, where $\sigma_{max}$ is the maximum traction (referred to here as the interfacial strength) and $\delta_f$ is the final separation (Fig. 5b). Under a given $W_{ad}$, the interfacial strength $\sigma_{max}$ is not uniquely determined when $\delta_f$ is varied. In energetic theories of adhesive contact (e.g., JKR theory), $W_{ad}$ alone is sufficient for describing adhesion. However, this is no longer the case when $\delta_f$ is comparable to the length scale of the contact.[14,44] Based on this line of reasoning, combined with the results in Fig. 4, we hypothesize that the interfacial strength $\sigma_{max}$ is the main governing parameter for fold nucleation rather than the typical $W_{ad}$.

To verify this hypothesis, we use the FEA model to examine three different cases of controlling adhesion: (1) increasing $W_{ad}$ by increasing $\sigma_{max}$ and holding $\delta_f$ constant (Fig. 5c); (2) increasing $\sigma_{max}$ while holding $W_{ad}$ constant and decreasing $\delta_f$ (Fig. 5d); and (3) increasing $W_{ad}$ by increasing $\delta_f$ and holding $\sigma_{max}$ constant (Fig. 5e-5f). The simulation results presented in Fig. 3 are based on Case 1, where $W_{ad}$ was assumed to increases proportionally with $\sigma_{max}$. Since the fold formation itself is rate-insensitive (Fig. 2d), we take the substrate to be a neo-Hookean solid for simplicity. In Fig. 5c, we show that in Case 1, the fold size increases from zero to non-zero as $\sigma_{max}$ is increased, indicating a transition from non-folding to folding (see snapshots in Fig. 5c). Interestingly, the $F_{peak}$ data exhibits a dip as the fold size increases, which is qualitatively similar to the experimental data in Fig. 4a. In Case 2 (Fig. 5d), we also find a non-folding to folding transition and a dip in $F_{peak}$ as $\sigma_{max}$ is increased, despite $W_{ad}$ being fixed. In contrast, when $\sigma_{max}$ is fixed (Case 3), a non-folding to folding transition is not observed, even though $W_{ad}$ is varied over an even larger range than Case 1 (Fig. 5e). To also test the effect of $\sigma_{max}$, in Fig. 5f we introduce a higher value $\sigma_{max}$, which is still held constant while increasing $W_{ad}$ (Case 3).

Although a non-folding to folding transition is not observed, the folds themselves appear (Fig. 5f). Explicitly, our results show that the substrate always exhibits non-folding when $\sigma_{max}$ is low (Fig. 5e) and always exhibits folding when $\sigma_{max}$ is high, regardless of whether $W_{ad}$ is fixed or varied. Combining the results from these four cases, we conclude that the interfacial strength $\sigma_{max}$ is the governing parameter for fold formation, rather than the adhesion energy $W_{ad}$; this helps resolve the seeming paradox observed in Fig. 4.

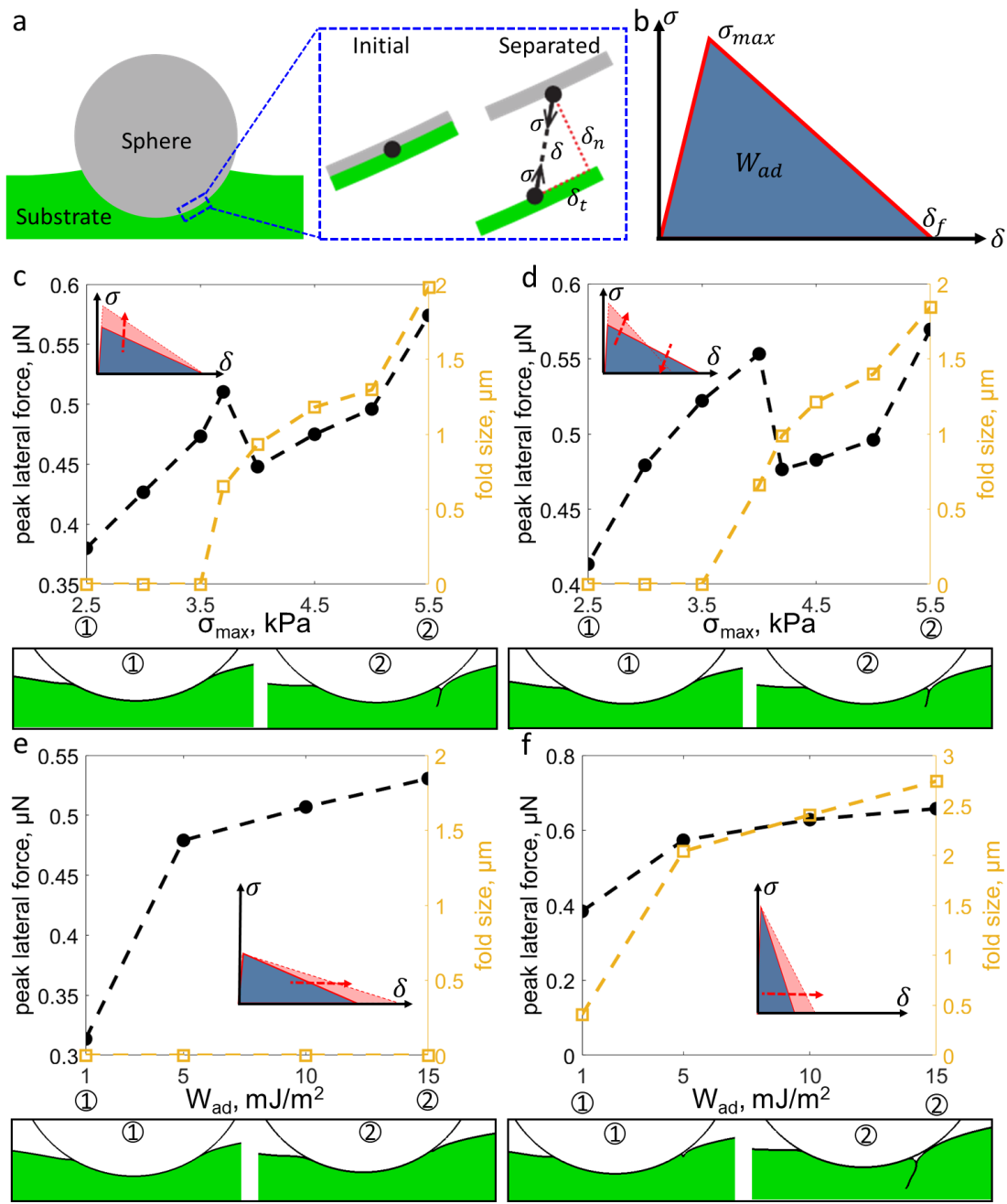

**Figure 5. Interfacial strength, not adhesion energy, governs fold formation.** (a) Schematic illustration of the contact between the rigid indenter and the substrate (left). The inset on the right shows a zoomed-in view of the interface in its initial state when contact is just established, and the separated state where the two surfaces are relatively displaced and thus a traction $\sigma$ and separation $\delta$ are used to define the interfacial adhesion. $\delta_n$ and $\delta_t$ are the normal and tangential separations, respectively. (b) A cohesive zone model with bilinear traction-separation law is used to define the adhesion energy $W_{ad}$ on the interface. (c) Top: Peak lateral force and fold size as a function of maximum strength $\sigma_{max}$ for Case 1. The inset shows the change in the cohesive zone model where $W_{ad}$ is increased by increasing $\sigma_{max}$ and holding $\delta_f$ = 2 μm constant. Bottom: Morphology of the system at the peak lateral force with $\sigma_{max}$ = 2.5 or 5.5 kPa. (d) Top: Peak lateral force and fold size as a function of $\sigma_{max}$ for Case 2. The inset shows the change in the cohesive zone model where $W_{ad}$ is fixed at 5 mJ/m² by increasing $\sigma_{max}$ and decreasing $\delta_f$. Bottom: Morphology of the system at the peak lateral force with $\sigma_{max}$ = 2.5 or 5.5 kPa. (e) Top: Peak lateral force and fold size as a function of $W_{ad}$ for Case 3. The inset shows the change in the cohesive zone model where $W_{ad}$ is increased by increasing $\delta_f$ and holding $\sigma_{max}$= 3 kPa constant. Bottom: Morphology of the system at the peak lateral force with $W_{ad}$ = 1 or 15 mJ/m². (f) Top: Peak lateral force and fold size as a function of $W_{ad}$ for Case 3. The inset shows the change in the cohesive zone model where $W_{ad}$ is increased by increasing $\delta_f$ and holding $\sigma_{max}$ = 6 kPa constant. Bottom: morphology of the system at the peak lateral force under $W_{ad}$ = 1 or 15 mJ/m².

*Discussion*

The folds in our study are reminiscent of both surface creases and Schallamach waves. Therefore, it is instructive to consider the similarities and distinctions of our folds to these phenomena. Self-contacting surface creases occur when a global compression is applied to a surface; for example, creases arise when attaching a soft layer to a pre-strained surface and releasing,[45] or simply applying a compressive strain on a soft block.[46,47] Through calculations, crease nucleation has been illustrated to be similar to a first-order transition,[47] and also sensitive to local imperfections on the surface.[48] Although confocal microscopy images of surface creases resemble our folds, they have a different curvature at the surface.[49] Moreover, near the elastocapillary length, defined as a ratio of surface tension and elastic modulus, $L_{EC} = \gamma/E$, creases can leave scars after removal of strain due to capillarity and self-adhesion. In our case, $L_{EC} \approx 4$ μm, assuming $\gamma = 20$ mN/m and $E = 5$ kPa. Yet no scars are observed once the self-contacting fold moves through the contact zone of the microparticle. The lack of scaring may be due to low adhesion hysteresis but, more likely, is due to the state of tension at the trailing edge that pulls the fold open. However, it is interesting to note that the dip in $F_{peak}$ (Fig. 4) occurs when the fold size is near $L_{EC}$. Our folds are similar to Schallamach waves, where a surface instability manifests from a build-up of compressive stress at the front, which then passes through the contact zone.[50] In contrast to conventional Schallamach waves however, our folds are self-contacting such that no air gap is visible. The fold maintains a self-contacting profile until it reaches the trailing edge and opens from the local tension. Additionally, the fold geometry is relatively stable until it is near the trailing edge; we can stop the probe for several minutes without the fold moving (Fig. 2c-2e). Hence, the folds discovered here bridge the physical mechanisms of surface creases and Schallamach waves.

From an experimental standpoint, our investigation reveals the significance of silicone aging on the interfacial mechanics of soft contacts. Soft materials, such as lightly crosslinked PDMS, often have uncrosslinked free chains that are not tethered to the polymer network.[37,51] We previously found that the PDMS used here (60:1 Sylgard 184) contains nearly 60% extractable materials.[38] These mobile molecules have been shown to form a liquid meniscus at the contact line of a stationary drop or microparticle.[35,51] In our case of a moving microparticle, the free chains are expected to reduce $\sigma_{max}$ by acting as a partial lubricant, lowering but not eliminating the network-particle contact.[34,52-56] At the same time, free chains can aid in maintaining high normal adhesion due to capillarity;[57] we associate this case to the freshly-prepared, untreated surfaces. However, minor stick-slip behavior may result from a mixed contact interface, where the particle is partially in contact with network and partially in contact with free chains (free mobile molecules) or dangling ends, which may also be the case for hydrogels.[2,34,54,58-60] After ambient light-aging (Fig. 2a), a line of increased fluorescence is observed at the surface. We also notice that the fluorescent signal of our fluorescent monomer increases after crosslinking into the network (Supplementary Note 4, Supplementary Fig. 6). Therefore, the increased fluorescence suggests some extra crosslinking at the surface. This additional crosslinking would reduce the amount of uncrosslinked free chains and dangling free ends near the surface that could partially lubricate the contact. Although this crosslinking may lead to a marginal increase in modulus at the surface, a stiff layer is not required for fold formation. We confirmed this by running a simulation with a stiff layer, which illustrates that a higher $W_{ad}$ is required to fold with an increasing stiff layer (Supplementary Note 4). Hence, we believe that the light-aged samples have a higher $\sigma_{max}$ due to a decrease in free chains and/or dangling ends near the surface.

In summary, we have introduced a unique approach to experimentally investigate the mechanism of static friction for a microscopic contact on a soft, adhesive surface. Using lateral force and confocal microscopy, we show that both folding and non-folding cases can occur. Folding emerges with sufficiently high interfacial strength between the particle and the surface, which manifests experimentally through light-aging or UVO exposure. Interestingly, our results show that a sufficiently high lateral resistance is required for folds to occur, although the folds themselves can lower the measured peak lateral force (i.e. a dip occurs in the $F_{peak}$ as a function of $\sigma_{max}$). By tuning the parameters that govern adhesion energy in simulations, we conclude that the maximum interfacial traction plays a dominating role in the fold-force relationship. Often, adhesive interactions are bucketed into an adhesive energy, $W_{ad}$. However, our results indicate that the interfacial strength $\sigma_{max}$, rather than $W_{ad}$, is a better predictor of the folding behavior. Hence, one key insight arising from our study is that $W_{ad}$ alone is not sufficient to predict the nucleation of folds or to predict $F_{peak}$, but rather the specific parameters that define $W_{ad}$ must be considered. Such results are anticipated to be important for soft adhesives, tribological surfaces, and self-cleaning coatings.

**Methods**

*Materials.*

Dow Sylgard 184 was purchased from Ellsworth Adhesives as a two-part kit. Polydisperse soda lime glass microspheres (2.5 g/cc) were purchased from Cospheric LLC. No. 1 glass coverslips and chloroform were purchased from VWR, and fluorescein diacrylate was purchased from Sigma-Aldrich.

*PDMS (polydimethylsiloxane) preparation.*

Sylgard 184, a commercially available polydimethylsiloxane two-part kit, is dyed with fluorescein diacrylate. The two parts are mixed with a 60 to 1 ratio, giving a modulus of 3.5 ± 0.5 kPa. The material was prepared following a previously described procedure. Briefly, fluorescein diacrylate is first dissolved in chloroform, mixed with Sylgard 184 base, and left for the chloroform to evaporate at 65°C. The curing agent was subsequently mixed with the dyed base.[34] The mixture was degassed under vacuum to remove trapped air for ~30 minutes, and spin-coated on a glass coverslip at 1000 RPM for 60 seconds to achieve a thickness of ~75 μm. Other RPMs were used to increase or decrease the thickness during control experiments to test the effect of sample thickness. [34] An RPM of 1000 was chosen to maximize the thickness of the PDMS while maintaining the necessary resolution using an optically correctable objective on the confocal microscope. After spin-coating, the coverslip with the uncured PDMS is placed in an oven at 65°C for 48 hours to cure.

*Characterization.*

*Imaging via confocal microscopy.* Images were taken of the fluorescein diacrylate dyed samples using an inverted Leica confocal microscope equipped with a 40x air objective. A correction ring on the objective enabled focusing on the air-PDMS interface through the thickness of the PDMS sample. Variable imaging rates and resolutions were used to optimize the imaging rate and quality. An excitation laser with a wavelength of 488 nm and a collection range of 495 to 520 nm was consistently used. Note that the laser power is left constant when comparing the fluorescence signal before and after crosslinking.

*Image analysis.* The confocal images were analyzed using ImageJ. A sphere was fit to the shape of the colloidal probe in the image to determine the location. To record the indentation depth, the distance from the lowest point of the sphere to the PDMS surface far outside the contact zone was measured, while the particle was at rest.

*Force microscopy.* A JPK Nanowizard 4 is mounted on top of the confocal microscope. A colloidal probe AFM cantilever was used for all measurements. A ~17 μm diameter glass sphere was attached to a thermal noise calibrated ACL-TL tipless cantilever having a nominal spring constant of ~40 N/m using high strength epoxy. For lateral force tests, the probe was indented into the PDMS at a rate of 1 μm/s and held for 15-20 seconds. To conduct lateral force measurements, a high-precision linear stage (Physik Instrumente, L-509), mounted to the side of the confocal microscope, was connected to a custom sample holder between the AFM head and the confocal microscope. This setup enables lateral translation of the sample, while the confocal microscope objective and the colloidal probe of the AFM remain aligned and in focus. Prior to these experiments, lateral force calibration of the cantilever was performed by scanning the tip across a clean glass slide at different normal loads. Based on a procedure in the literature, we assumed a coefficient of friction of 0.4, and calculated the frictional force relative to the cantilever deflection.[61] For normal adhesion tests, the probe was indented into the surface at a rate of 0.1 μm/s to a depth of ~0.2 μm, followed by retraction at the same rate. To test if a folding behaviour is affected by the dye itself, we conducted experiments without dye but in reflection mode on the confocal microscope. Although this imaging is not able to measure details, fold release can still be observed (Supplementary videos 3 and 4).

*Simulations*

Finite Element Analysis (FEA) was performed to simulate the microscale indentation and sliding experiments using the commercial package ABAQUS (version 2020, Simulia, Providence, RI, USA). The FEA model consisted of two components: a rigid sphere and a viscoelastic substrate. To reduce computational cost, a two-dimensional (2D) plane strain model was built where the spherical indenter was modelled as a rigid circle with diameter of 17 μm, and the substrate was a 200 μm × 50 μm rectangle meshed with 2D plane strain elements (CPE4RH). Adhesion between the indenter and the substrate was modelled by a cohesive zone following the bilinear traction-separation law. More information on the material model, cohesive zone and mesh convergence test is provided in Supplementary Note 2. Initially the indenter was on top of the substrate surface with a 1 μm gap between the indenter and the substrate surface. In a simulation, the indenter was firstly moved downwards at a velocity of 1 μm/s until a desired indentation depth was achieved. After a 17-second relaxation step (to account for the experimental time gap between normal indentation and lateral motion), the indenter was moved horizontally with the indentation depth held fixed. All simulations were performed using dynamic/implicit solver to accommodate the instability associated with folding. Since physically the 2D plane strain model represent the cross-section of an infinitely long cylinder in contact with the substrate, the lateral force obtained from the simulations is a line force, i.e., force per unit length along the out-of-plane direction. To enable comparison with experimental data, the total lateral force was defined by multiplying the line force by the indenter diameter of 17 μm.

**Acknowledgements**

The authors acknowledge support from the National Science Foundation; J.D.G. and J.T.P through CMMI-1825258 and X.Y. and R.L through a CAREER award, CMMI-1752449.

**Supplementary information**

**Supplementary Note 1: Control experiments for folding**

As discussed in the main text, we found that samples left to age in ambient light exhibited folding behavior. This supplementary note describes a range of different experimental controls used to conclude that folds appear after light-aging. These variables include the Sylgard 184 mixing ratio (modulus), particle size, varying dwell times, sample thickness, and extended curing times; however, folding was consistently on a per sample basis. In other words, light-aged samples always exhibited folds while freshly prepared samples did not fold. To test if slight modulus variations, which can occur in batch-to-batch sample preparation, caused folding, we prepared samples with both 50 and 70 to 1 mixing ratios of Sylgard 184; these possess slightly higher and lower modulus respectively. However, no folds were observed on the freshly prepared samples. Hence, we kept our samples constant at the 60:1 mixing ratio. Two probe sizes were also tested including $R$=8.5 and 13.5 µm; again, no folds were observed on freshly prepared samples. To ensure that the folding is not due to slight variations in dwell time, we varied the dwell time from 0 to 30 seconds; folding was still not observed on fresh samples. The sample thickness of the PDMS substrate was also confirmed to not be the cause of folding, since both folding and non-folding behavior was observed on samples of the same thickness from ~30 µm to ~90 µm. To consider if extended cure times lead to folding, we cured samples from 24 hours to one week at 65 °C. After one week in a dark oven, no folds were observed. After conducting these control experiments, we find that freshly prepared samples never fold due to slight modulus variations, sample thickness, particle size, curing time, and dwell time (within the ranges tested).

**Supplementary Note 2: Details for Finite Element Analysis (FEA)**

All simulation results presented in the main text were generated using a two-dimensional (2D) plane strain model consisting of a rigid indenter and a deformable substrate. The simulation steps were adopted according to experimental procedures, as summarized in the Methods section of the main text. This supplementary note elaborates three aspects of the FEA model: i) the material model for the deformable substrate, ii) the cohesive zone model for the adhesion between indenter and substrate, and iii) the mesh convergence tests.

*Material model*

The PDMS substrate was modelled as an incompressible visco-hyperelastic model that combines the neo-Hookean solid and a Prony series to capture viscoelasticity under finite deformation. This model, based on the framework of Simo,[1] extends the formulation of linear viscoelasticity to accommodate finite deformation kinematics and hyperelasticity, and is readily available in ABAQUS (version 2020, Simulia, Providence, RI, USA). Since the incompressible neo-Hookean solid is characterized by only the shear modulus, the corresponding visco-hyperelastic model can be fully specified by the relaxation function, which is given below according to the Prony series:

$$G(t) = G_\infty + \sum_{i=1}^{N} G_i \exp(-t/\tau_i), \qquad (S1)$$

where $G_\infty$ is the long-term shear modulus in the fully relaxed limit, $N$ is the number of relaxation modes, $G_i$ and $\tau_i$ are the shear modulus and relaxation time associated with each relaxation mode, respectively. In the frequency domain, Eq. (S1) results in the following functions of the storage shear modulus $G'$ and the loss shear modulus $G''$:[2]

$$G'(\omega) = G_\infty + \sum_{i=1}^{N} G_i \frac{\omega^2 \tau_i^2}{1+\omega^2 \tau_i^2}, \qquad (S2)$$

$$G''(\omega) = \sum_{i=1}^{N} G_i \frac{\omega \tau_i}{1+\omega^2 \tau_i^2}, \qquad (S3)$$

where $\omega$ is the angular frequency (unit: rad/s). The parameters $G_\infty$, $G_i$ and $\tau_i$ ($i$ = 1, 2, ... N) were calibrated by fitting rheology data of the PDMS substrate using Eq. (S2) and Eq. (S3). Supplementary Fig. 1 shows the fitted curves together with the experimental data. The long-term shear modulus $G_\infty$ is found to be 1.89 kPa and the other fitting parameters are listed in Table 1. Under extremely low lateral velocity ($v \to 0$), the viscoelastic substrate is expected to be in its long-term relaxed limit. We performed simulations for pure elastic substrate to study this limit by simply removing the relaxation function, which recovered an incompressible neo-Hookean solid with a shear modulus of 1.89 kPa for the substrate.

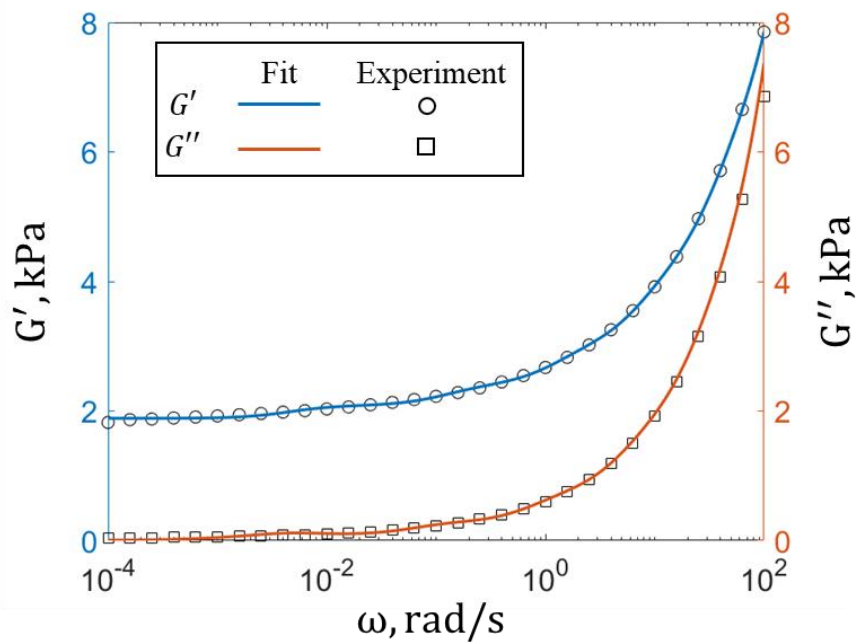

**Supplementary Figure 1:** Storage ($G'$) and loss ($G''$) shear moduli as a function of the angular frequency $\omega$. Symbols represent experimental data (circle: $G'$; square: $G''$) from rheological tests of the PDMS substrate. Solid lines represent fits (blue: $G'$; orange: $G''$) of the Prony series. The rheological data were obtained by shear rheology using a ~1 mm thick, 60 to 1 Sylgard 184 sample between 25 mm parallel plates in the linear regime.[3]

**Supplementary Table 1:** Calibrated parameters for the relaxation in Eq. (S1). Note that $G_\infty$ = 1.89 kPa.

| $i$ | 1 | 2 | 3 | 4 | 5 | 6 |
|---|---|---|---|---|---|---|
| $G_i$ (kPa) | 31.3 | 2.65 | 0.324 | 1.207 | 0.2 | 0.617 |
| $\tau_i$ (s) | 0.0021 | 0.249 | 8.0712 | 0.1264 | 221.923 | 0.7953 |

*Cohesive zone*

Adhesion between the indenter and the substrate was captured using a cohesive zone mode. This model prescribes a traction-separation relation between the two interfaces,[4] as schematically illustrated in Supplementary Fig. 2a. Briefly, when two contacting points on the indenter and the substrate are separated by a vector of δ, they are subjected to an attractive traction σ (force per unit area) induced by adhesion. Both vectors of σ and δ can be resolved to components along the normal and tangential directions of the interface. Note that a 3D model has two tangential directions while a 2D model only has one tangential direction. We assumed an uncoupled and isotropic traction-separation relation, implying that σ and δ are along the same direction. Therefore, the traction-separation law can be specified by relating the magnitudes of σ and δ, which are denoted as $\sigma$ and $\delta$, respectively. In general, it is challenging to directly measure the relation between $\sigma$ and $\delta$. To capture the essential physics, we adopted a simple bilinear traction-separation relation featuring three parameters: the maximum separation $\delta_\mathrm{f}$, the interfacial strength $\sigma_\mathrm{max}$, and the initial stiffness $K$. Alternatively, the adhesion energy $W_\mathrm{ad} = \sigma_\mathrm{max}\delta_\mathrm{f}/2$ (i.e., the area underneath the traction-separation curve) is a parameter representing energy required to separate a unit area of the interface. As shown in Supplementary Fig. 2b, $\sigma$ first increases linearly with $\delta$ with a slope of $K$. Interface damage is initiated when the interfacial strength $\sigma_\mathrm{max}$ is achieved. Specifically, we used the maximum stress criterion for damage initiation, which is stated as:

$$\max\left\{\langle\sigma_n\rangle, \sigma_s, \sigma_t\right\} = \sigma_{\max}, \tag{S4}$$

where $\sigma_n$, $\sigma_s$ and $\sigma_t$ represent the normal traction component and the two tangential traction components, respectively, and the Macaulay bracket $\langle \bullet \rangle$ is to signify that a compressive stress does not initiate damage. Complete interface failure occurs when $\delta_f$ is reached and the traction reduces to 0. In this work, we used $\delta_f = 2$ µm and $K = 2 \times 10^{11}$ N/m$^3$ if otherwise specified. The values of $\sigma_{max}$ and $W_{ad}$ were varied to probe the effects of interfacial strength and adhesion energy.

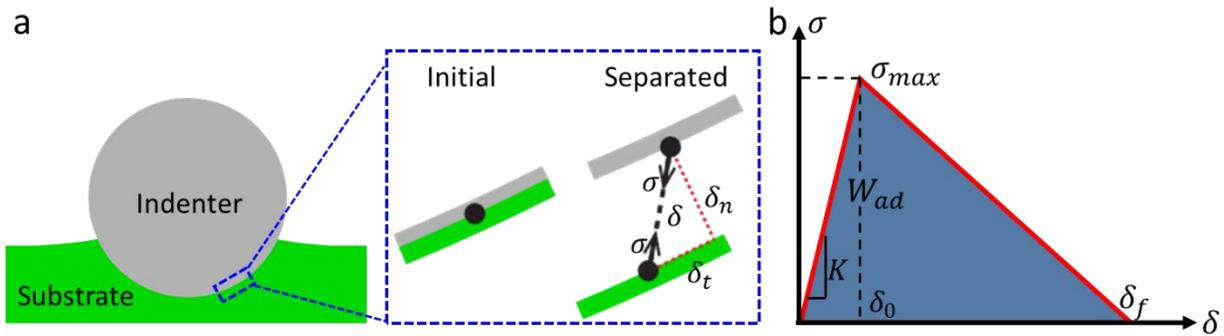

**Supplementary Figure 2.** Cohesive zone model for the adhesive interface. (a) Schematic showing the contact interface between the indenter and the substrate (left). The inset on the right shows a zommed-in view of the interface in its initial state when contact is just established, and the separated state where the surfaces are relatively displaced by $\delta$ and subjected to an attractive traction $\sigma$. $\delta_n$ and $\delta_t$ are the normal and tangential separations, respectively. (b) Illustration of the bilinear traction-separation relation for the cohesive zone model.

*Mesh Convergence*

The mesh of the FEA model is shown in Supplementary Fig. 3a. The substrate was meshed with 2D plane strain elements (CPE4RH) with uniform mesh size along the horizontal direction and biased mesh size along the vertical direction. The smaller elements were located at the surface of the substrate and were square in shape with an edge length of $l_m$. The indenter was modelled as 1D rigid wire with an element size of 0.1 µm. Recall that the diameter of the indenter was 17 µm. To test if the mesh is sufficiently fine to resolve the highly localized deformation associated with folding, we performed a mesh convergence test using a benchmark case where the substrate was elastic (i.e., incompressible neo-Hookean solid with shear modulus being 1.89 kPa) and was subjected to an indentation depth of $0.4R = 3.4$ µm ($R = 8.5$ µm is the indenter radius). The adhesion energy $W_{ad}$ was set to be 8 mJ/m² to enable folding. We carried out a series of simulations with the smallest substrate element size $l_m$ varying from 4 μm to 0.2 μm, and extracted the peak lateral force and fold size (Supplementary Fig. 3b) as a function of the mesh size $l_m$. As mentioned in the Methods section of the main text, the lateral force given by the 2D simulations is a line force (i.e., force per unit length along the out-of-plane direction) and was converted to the total lateral force by multiplying the line force by the indenter diameter. The fold size was defined as half of the contour length from one end of the fold to the other. The data in Supplementary Fig. 3c clearly shows that the simulation result, in terms of the peak lateral force and fold size, converges when $l_m$ is smaller than 0.4 μm. Therefore, we have adopted $l_m$ = 0.2 μm in all of our simulations.

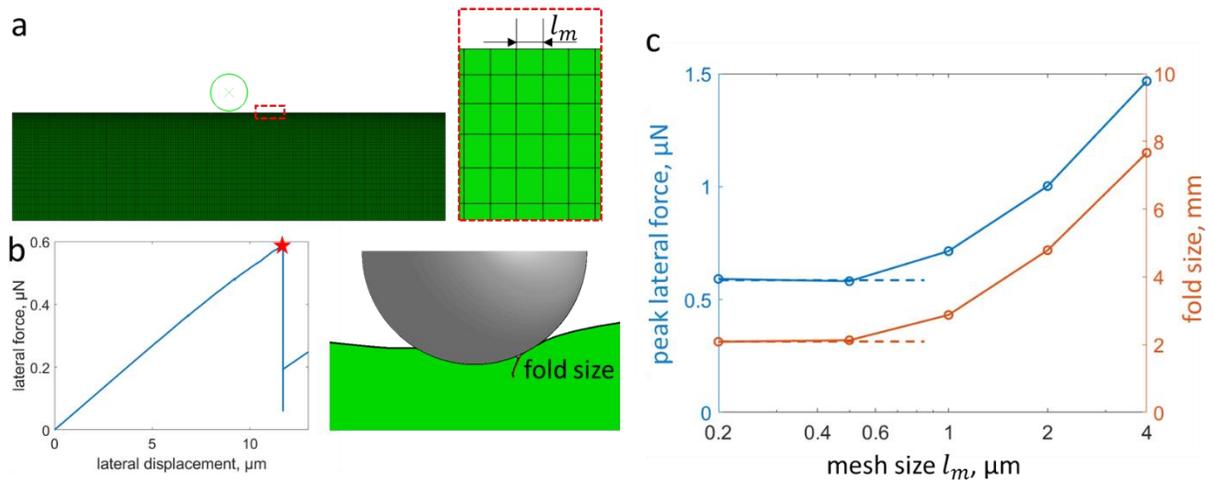

**Supplementary Figure 3.** Mesh convergence test. (a) Finite element mesh for the 2D model (left). The inset shows a zoomed-in view of the mesh near the substrate surface consisting of square elements with an edge length of $l_m$. (b) Peak lateral force (marked by the red star on the plot of lateral force versus lateral displacement) and fold size obtained from a representative simulation. (c) Peak lateral force and fold size as a function of the mesh size $l_m$ (log-scale). The results converge when $l_m$ is less than 0.5 μm.

**Supplementary Note 3: Discussions on the discrepancy between simulations and experiments**

There are quantitative discrepancies between the simulation results and experimental data, as shown by the comparison between Fig. 2 and Fig. 3 in the main text. Although the goal of our simulations is to obtain physical insights towards the underlying mechanism of fold formation, rather than to quantitatively reproduce the experimental observation, it is important to understand the sources of the discrepancies. This note discusses the sources of three main discrepancies: i) 2D versus 3D model; ii) lower peak lateral forces in simulations; iii) lateral force during sliding in the non-folding cases.

*2D versus 3D model*

The 2D plane strain model adopted in our simulations should be interpreted as the cross-section of an infinitely long cylinder in contact with the substrate, which is obviously different from the experimental geometry with a spherical indenter. To shed light on the effect of this difference, we built a 3D model as shown in Supplementary Fig. 4a. The substrate was meshed by linear brick elements (C3D8RH). The smallest elements are located at the center of the top surface with an element size of 0.2 μm × 0.2 μm × 0.2 μm, which is similar to the 2D model. The indenter was modeled as a rigid shell surface with an element size of 0.2 μm × 0.2 μm. All other simulation parameters (e.g., material model, cohesive zone model and analysis steps) are identical to those of the 2D model.

Supplementary Fig. 4b-4d compare results of the 2D and 3D models with indentation depth = $0.4R$ (i.e., 3.4 μm), velocity = 1 μm/s, and $W_{ad}$ = 20 mJ/m². Folding is observed in both 2D and 3D simulation results. Unlike the 2D model where the fold is straight along the out-of-plane direction (Supplementary Fig.4b), in the 3D model the fold is curved: it follows the circular contact perimeter between the indenter and the substrate at the leading edge and extends towards the side of the contact region, which is consistent with the experimental observation (Fig. 1c of the main text). Nevertheless, the fold shape on the central vertical cross-section of the 3D model is similar to that of the 2D model. In addition, the 2D and 3D models give a similar trend of lateral force (Supplementary Fig. 4d). Recall that the lateral force in the 2D model is a line force (i.e., force per unit length along the out-of-plane direction). To enable comparison with the 3D model,

we have converted the 2D line force to the total lateral force by multiplying the line force by the indenter diameter (17 μm). After conversion, we find the peak lateral forces in 3D and 2D models are close. However, the 2D model predicts a higher compliance than the 3D model, i.e., it takes a longer lateral displacement to reach the peak lateral force in the 2D model. We attribute this difference to 3D effects of the fold morphology. Note that the formation of the fold increases the compliance for lateral motion of the indenter. In the 3D model, the fold size is the largest at the central cross-section and decreases towards the two sides. In contrast, in the 2D model the fold size is uniformly large along the out-of-plane direction, thereby leading to a larger increase in the compliance. Despite the difference in compliance, the comparison in Supplementary Fig. 4 shows that the 2D model can capture the essential physics of fold formation. The 3D model contains 20~30 times more elements than the 2D model with the same mesh density, and thus is much more computationally expensive. Therefore, we adopted the 2D model for the parametric study on fold formation.

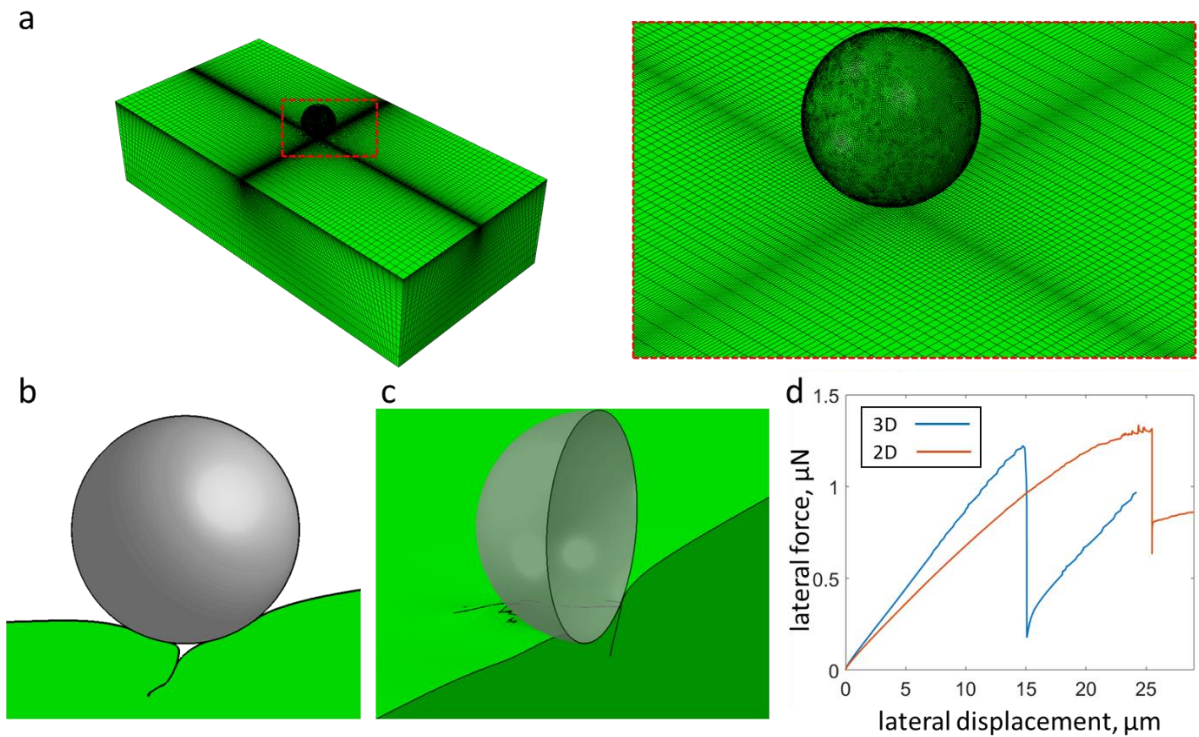

**Supplementary Figure 4.** Comparison between the 3D and 2D model under indentation depth = 0.4$R$, lateral motion velocity = 1 μm/s, and $W_{ad}$ = 20 mJ/m². (a) Geometry and mesh of the 3D model (left) and a zoomed-in view of the region near the indenter (right). (b) Snapshot of the 2D model at the peak lateral force. (c) Snapshot of the 3D model at the peak lateral force. (d) Lateral force versus the lateral displacement. For the 2D model, the line force is multiplied by the indenter diameter to enable comparison with the lateral force obtained from the 3D model.

*Lower peak lateral forces in simulations*

Although the simulation results in Fig. 3 of the main text capture the qualitative trend of the experimental data (Fig. 2 of the main text), the peak lateral force obtained from simulations are consistently smaller than the experimental data. Such difference cannot be explained by the 2D approximation adopted in the simulations, given that the 2D and 3D models give similar peak lateral force (Supplementary Fig. 4d). Instead, we attribute the lower peak lateral forces in simulations to the cohesive zone parameters. As described in Supplementary Note 2, the cohesive zone model features a bilinear traction-separation relation with several parameters, e.g., $\sigma_{max}$, $\delta_f$, $K$, and $W_{ad}$. The peak lateral force may be dependent on each of these parameters, especially when folding occurs. For example, Fig. 5d in the main text shows that the peak lateral force and fold size can increase with $\sigma_{max}$, even if $W_{ad}$ is held fixed. Since direct measurement of these cohesive parameters is difficult, if not impossible, a multi-variable optimization process is required to determine their values by matching the simulation results to experimental data. This is not pursued in this work since our focus is on the physical mechanism of ford formation. In addition, using smaller $\delta_f$ or larger $\sigma_{max}$ would require a finer mesh to achieve convergence, which can significantly increase the computational cost.

*Lateral force during sliding in the non-folding cases*

For the non-folding case, the simulation result (Fig. 3c in main text) exhibits a substantial drop in the lateral force after initial sliding starts, whereas experimental data (Fig. 2c in main text) shows that the lateral force during initial sliding remains at a similar level to the peak force. This is because the cohesive zone model in the simulation is irreversible. Briefly, cohesive interaction is activated when elements on the substrate surface first get in contact with the indenter surface at the leading edge of the indenter. Once the cohesive interaction fails, elements on the substrate surface cannot re-establish the cohesive interaction with the indenter element. Therefore, once sliding starts, the cohesive interaction is limited to only a few elements at the leading edge that are newly in contact with the indenter, which results in a lower lateral force than before sliding.

To illustrate this point, we plotted the interface damage parameter along the contact region during sliding for the non-folding case (Supplementary Fig. 5). The damage parameter describes degradation of cohesive interaction between two elements and ranges from 0 to 1. The cohesive interaction is fully intact when the damage parameter is equal to 0, and totally fails when it becomes 1. Supplementary Fig. 5b shows that during sliding, most of the contact region exhibits an interface damage parameter of 1, indicating no cohesive interaction. Only a few elements near the leading edge of the indenter exhibit a damage parameter less than 1, which contributes to the small lateral force during sliding. The irreversibility of conventional cohesive zone model is discussed in a recent work,[5] where a reversible cohesive zone model was developed. It should be emphasized that this artifact only affects the sliding stage and does not impact the static friction stage (i.e., before the peak lateral force) that is the focus on our work, where there is no need to re-establish cohesive interactions.

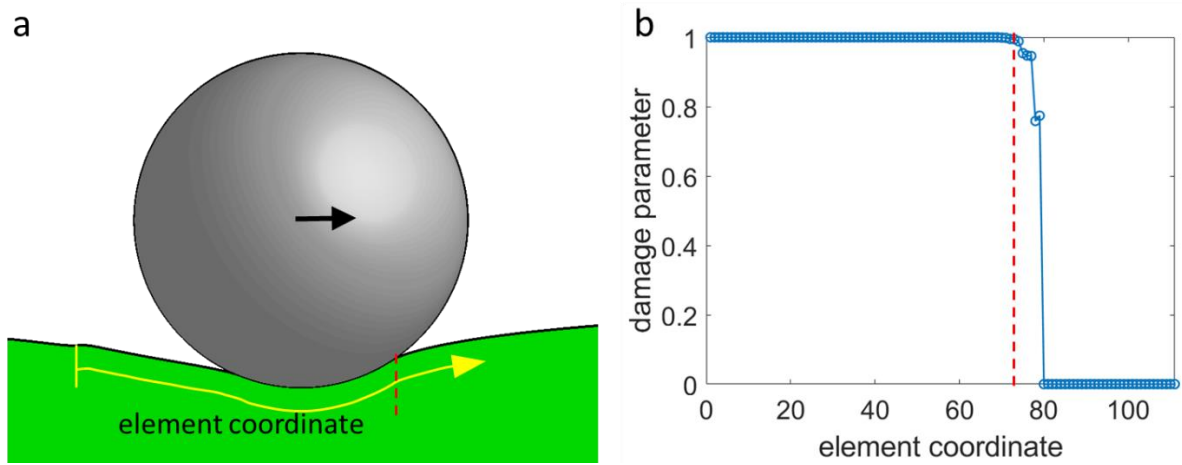

**Supplementary Figure 5.** Interface damage parameter during the sliding stage of a non-folding case with indentation depth = 0.4$R$, lateral velocity = 1 μm/s, and $W_{ad}$ = 4 mJ/m². (a) Snapshot from the simulation in the sliding stage. The yellow arrow illustrates the range and direction of element

coordinate on the substrate surface. (b) Damage parameter as a function of element coordinate on the substrate surface. The red dashed line highlights the location where the damage parameter becomes less than 1, which is also marked on part (a).

**Supplementary Note 4. Fluorescence intensity with crosslinking and potential top layer effects**

In this supplementary note, we expand on the question of fluorescence intensity and potential top layer effects. As described in the discussion of the main text, we observed a change in the fluorescent intensity of the crosslinkable, fluorescein diacrylate dye before and after curing (Supplementary Fig. 6). This suggests that a bright line at the PDMS surface may be the result of additional crosslinking (e.g. Fig. 2a in the main text). Additionally, UVO is known to create a glassy layer at the surface of PDMS. However, at our short times (tens of seconds), it is likely negligible in terms of the mechanics of folding (a UVO layer is likely less than ~1 nm).

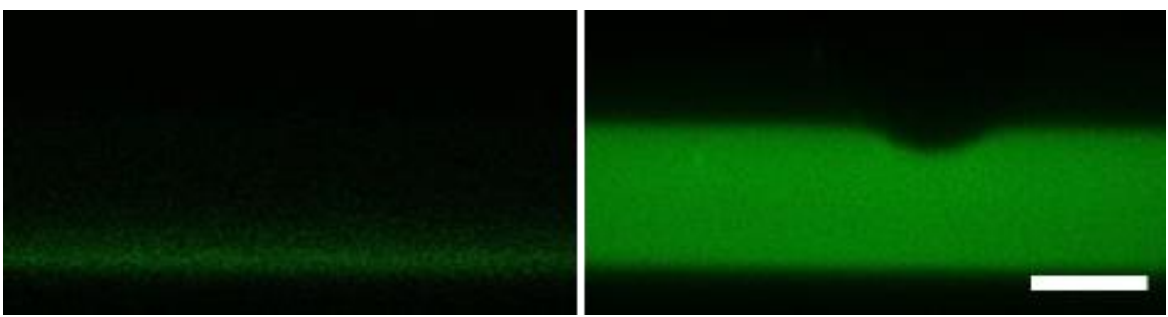

**Supplementary Figure 6.** Fluorescent signal of the Fluorescein diacrylate dye molecule before (left) and after curing (right). Note that the images are taken with the same microscope parameters (e.g. excitation laser power and collection wavelength). Scale bar: 20 µm.

To explore how a potential stiff layer might affect fold formation, we ran simulations with the 2D model to study the effect of stiff layer. In these simulations, the substrate was modelled as an incompressible neo-Hookean solid with a shear modulus of 1.89 kPa (i.e., neglecting viscoelasticity). We added a thin layer which is also modelled as an incompressible neo-Hookean solid but with a shear modulus that is 100 times larger than the substrate. The thickness of this stiff layer was varied from 50 nm up to 400 nm. The stiff layer was meshed by squared elements with a size less than 200 nm (recall that the smallest element size in the substrate was also 200 nm). Specifically, for 400 nm-thick layer, the mesh size in the layer was 200 nm × 200 nm. For 50 nm-thick layer, we refined the mesh such that the mesh size in the layer was 50 nm × 50 nm. To improve accuracy, the element type in these simulations were changed to CPE4H. The simulation results on the effect of the stiff thin layer are summarized in Supplementary Fig. 7. Specifically,

Supplementary Fig. 7a plots the minimum adhesion energy $W_{ad}$ required for fold formation as a function of the thin layer thickness, and Supplementary Fig. 7b shows the peak lateral forces corresponding to the data in Supplementary Fig.7a. Interestingly, when the stiff layer becomes thicker, a higher adhesion energy is required for fold formation. In addition, with a thicker stiff layer, the fold becomes larger and exhibits a rounded tip as opposed to the crease-like shape observed when the stiff layer is thin (see the inset of Supplementary Fig.7a). These results are consistent with the expectation that the stiff layer is associated with a bending stiffness, which brings an extra energy penalty to surface folding. Consequently, higher adhesion energy is required for fold formation with a thicker stiff layer; the peak lateral force also increases with thickness correspondingly. Hence, it is unlikely that a ~1 nm layer in our experiments, created by short UVO exposure, will play a significant role in fold formation.

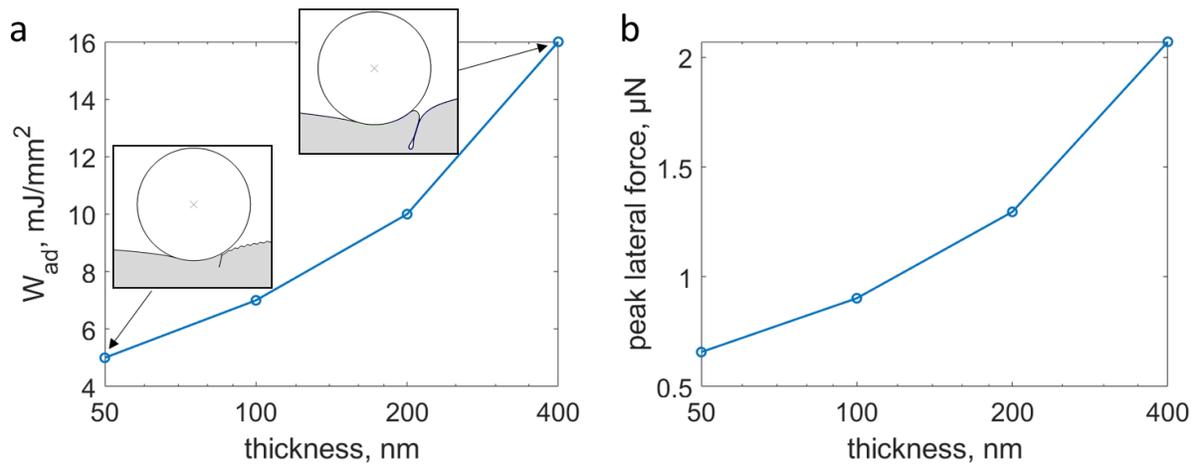

**Supplementary Figure 7.** Simulating effect of the stiff layer on an elastic substrate with indentation depth = 0.4 $R$. (a) Minimum adhesion energy required for fold formation as a function of the stiff layer thickness. The insets are the simulation snapshots at peak lateral force for the two cases with layer thickness being 50 nm or 400 nm. The other cohesive parameters are the same as those described in Supplementary Note 2 (i.e., $\delta_f = 2$ μm and $K = 2 \times 10^{11}$ N/$m^3$). (B) Peak lateral force corresponding to the data in part (a).